\documentstyle[12pt]{article}
\begin{document}
\begin{titlepage}
\renewcommand{\thefootnote}{\fnsymbol{footnote}}
\vspace {0.4cm}
\begin{center} {\LARGE\bf Ultraviolet-Renormalon Reexamined}
\end{center}
\vspace {0.4cm}
\begin{center} {\Large A.I. Vainshtein}\footnote{E-mail:
vainshte@vx.cis.umn.edu}\\
{\it Theoretical Physics Institute, University of Minnesota,\\
116 Church St. SE, Minneapolis, MN 55455}\\
and\\
{\it Budker Institute of Nuclear Physics, 630090 Novosibirsk,
Russia}\\
\vspace {0.4cm}
{\Large V.I. Zakharov}\footnote{On leave of absence from
Max-Planck-Institute of Physics, Munich.}\\
{\it University of Michigan, Ann Arbor, MI 48109}
\vspace {0.7cm}
\end{center}
\begin{center} {\Large\bf Abstract}
\end{center}
We consider large-order perturbative expansions in QED and QCD.
The coefficients of the expansions are known to be dominated by
the so called ultraviolet (UV) renormalons which arise from inserting
a chain of vacuum-polarization graphs into photonic (gluonic) lines.
In large orders the contribution is associated with virtual momenta
$k^2$ of order $Q^2e^n$ where $Q$ is external momentum, $e$ is the base of
natural logs and $n$ is the order of perturbation theory considered.
To evaluate the UV renormalon we develop formalism of operator product
expansion (OPE) which utilizes the observation that $k^2\gg Q^2$.
When applied to the simplest graphs the formalism reproduces the known
results in a compact form. In more generality, the formalism reveals
the fact that the class of the renormalon-type graphs is not well defined.
In particular, graphs with extra vacuum-polarization chains are not
suppressed. The reason is that while inclusion of extra chains
lowers the power of $\ln k^2$ their contribution is enhanced by
combinatorial factors.

\vspace{1cm}

\end{titlepage}

\newpage
\addtocounter{footnote}{-2}
\section{Introduction}

Large-order behavior of perturbative expansions has been studied for
about forty years starting from the the seminal paper by
Dyson \cite{Dyson} (for review and further references see Refs.
\cite{Book,Mueller1}). Although a lot of insight has been gained
some practical aspects of the issue in case
of QCD have not been clarified so far and attracted for this reason a
renewed attention recently. In particular, let us mention here
discussion of possible $1/s$ power corrections at large energies $\sqrt s$, as
generated through divergences of perturbative expansions
\cite{VZrecent,Brown,Beneke2,Maxwell}.
In this note \footnote{The letter-type version of the paper was
published in the Physical Review Letters, {\bf 73}, 1207 (1994)}
we address the problems of calculation
and of calculability of the so called ultraviolet renormalon
\cite{thooft,Lautrup,Parisimis,Bergere} which are related to these
$1/s$ corrections but of more general nature.

The importance of the ultraviolet renormalon rests
on the observation that it dominates large orders
both in QCD and QED \cite{thooft}. To be more specific we shall have
in mind perturbative calculations of the famous ratio $R$ :
\begin{equation}
R(s)~=~{{\sigma (e^+e^-\rightarrow
{}~hadrons)}\over
{\sigma(e^+e^-~\rightarrow~\mu^+\mu^-)}}
\label{1}
\end{equation}
as function of energy $\sqrt s$ and its Euclidean counter-part $\Pi(Q^2)$,
\begin{equation}
-Q^2{{d\Pi(Q^2)}\over {dQ^2}}~=~{{Q^2}\over{12\pi^2}}~\int_0^{\infty}
{{{R(s)~ds}\over{(s+Q^2)^2}}}
\label{2}
\end{equation}
as function of Euclidean momentum $Q$.
The quantity $Q^2d\Pi(Q^2)/dQ^2$ is represented as an expansion
in QCD coupling constant $\alpha_s(Q^2)$ :
\begin{equation}
-Q^2{{d\Pi(Q^2)}\over {dQ^2}}~=~(parton~model)\times \sum_{n=0}^
{n=\infty}a_n\alpha^n_s(Q^2)~~,\label{first}
\end{equation}
where first three coefficients have been calculated explicitly \cite{pertu}.

Note that the normalization of the coupling at $Q^2$, as above,
guarantees that there are no log factors in the expansion
(\ref{first}). Since we consider $Q^2$ to be large the  coupling
$\alpha_s (Q^2)$ is small.
For the sake of simplicity we will illustrate our approach mostly
on example of $U(1)$ model where strong interactions are mediated by
an abelian vector field. Then we consider energy to be far below the
position of the Landau pole so that $\alpha_1$ is small. As we shall
see the conclusions are of general nature and apply both to the
$U(1)$ model and QCD.

Now, the generic behavior of the expansion coefficients
at large $n$ looks as
\begin{equation}
{a_n}~\stackrel{n \to \infty}{\longrightarrow}
K n^{\gamma}{{n!}\over{S^n}}
\label{factorial}
\end{equation}
where $K,\gamma$ and $S$ are constants and actually there exists
a variety of
sources for the factorial growth (\ref{factorial}) resulting in different
$S$. The constant $S$ takes the smallest absolute value in case of the
ultraviolet renormalon:
\begin{equation}
S_{UV~renorm}~=~-\frac{1}{b_0}
\label{UV}
\end{equation}
where $b_0$ is the first coefficient in the corresponding $\beta$ function:
\begin{equation}
Q^2{d\over{dQ^2}}\left(\alpha(Q^2)\right)
{}~=~-b_0\alpha^2(Q^2)~-~b_1\alpha^3(Q^2)~+...
\label{Q}
\end{equation}
Note that in QCD $b_0$ is positive and the series (\ref{factorial}) is
sign alternating while in $U(1)$ case $b_0$ is negative.
The sign alternation allows for the Borel summation of the series and
from this point of view there exists an important
difference between QCD and $U(1)$. In this paper we are concerned mostly
with calculating of $a_n$ at large $n$, not on the
summation of the perturbative series.

The main technical tool we are using is the operator product expansion
(OPE), or the expansion in inverse powers of $k^2$ where $k$
is the virtual momentum flowing through the gluonic line (see Fig.~1).
The idea of using this OPE is outlined first by Parisi \cite{Parisimis} and
is based on the observation that effectively $k^2 \gg Q^2$ in case of the
ultraviolet renormalon. More specifically, we shall see later that
\begin{equation}
(k^2)_{eff}~\sim Q^2 e^n.
\label{p}
\end{equation}
where $n$ is the order of perturbative expansion and $e$ is the base
of natural logs. We will elaborate the idea of the expansion in
$Q^2/k^2$ on explicit examples.

Our primary objective in this paper is to
work out a scheme for explicit evaluation of renormalon
contributions. In Sec. 2 we will address computation
of the simplest renormalon-type graphs represented
on Fig.~1. In $U(1)$ case such calculations have been performed in
a number of papers \cite{Co,Kawai,Beneke2,Broadhurst}.
In particular we reproduce
the results of Beneke \cite{Beneke2} who evaluated the contribution
of the renormalon chain directly, without use of OPE.
What we would like to add here is a simplification of the scheme.
Moreover, the generalization to the QCD case is imminent.
As the next step we generalize the procedure to the case of two renormalon
chains, i.e. three-loop skeleton graphs (Sec. 3). We will demonstrate that
this three-loop contribution dominates over the two-loop one.

The consideration of graphs with one or two renormalon chains naturally
brings us to the problem of calculability
of the renormalon contribution in general (Sec. 4).
By this we mean the problem of identifying
the graphs which control the coefficient
in front of the factorial (see Eq. (\ref{factorial})).
In particular, one may increase the number of renormalon chains
and the question is whether this contribution is suppressed or
not. \footnote{
In case of QCD because of the complexity of the gauge-fixing problems
even the set of the graphs delegated to each type
is not so well defined a priori. We shall see, however, that the technique
developed allows to circumvent this
problem as well.}

Superficially, the more complicated graphs can be neglected since
at least one extra factor of $\alpha$ is not accompanied in this case
by a log. The use of the operator product expansion
mentioned above allows to analyze the problem in a general way.
The result turns unexpected: the contribution
of the graphs with extra chains is {\it not} suppressed
and the class of the ``renormalon-type''
graphs is {\it not} well defined at all.

The reason is that we have in fact two
large parameters, namely, $\ln k^2/Q^2$ and the order of the perturbative
expansion, $n$. Moreover, effectively
$\ln k^2/Q^2 \sim n$, as is implied by Eq. (\ref{prop}).
While the coefficient in front of the
highest power of the log can be found in
a straightforward way -- and this is the
essence of the renormalization group analysis --
it turns to be {\it not} enough to evaluate the asymptotic of $a_n$.
The other
contributions lose the log factors but gain
extra factors of $n$ because of combinatorics.

\section{Operator product expansion. Two-loop example.}

Let us consider polarization operator $\Pi_{\mu\nu}$,
\begin{equation}
\Pi_{\mu\nu}~=~i\int{
dx e^{iqx}\langle 0|
T\{j_{\mu}(x)j_{\nu}(0)|0\rangle}=
(q_{\nu}q_{\nu}-g_{\mu\nu}q^2)\Pi(Q^2),~~~Q^2=-q^2\label{pp}
\end{equation}
of electromagnetic current $j_{\mu}$ in the simplified model with $N_f$
fermionic fields
\begin{equation}
j_{\mu}~=~\sum_{q}Q_q\bar q\gamma_{\mu}q\label{current}
\end{equation}
where $Q_q$ are the corresponding electric charges. Strong interactions are
mediated by $U(1)$ gluonic field $B_{\mu}$ and the Lagrangian of
the model is
\begin{equation}
L~=~-{1\over 4} B_{\mu\nu}B^{\mu\nu}+
\sum_{q}\bar q\gamma^{\mu}i\partial_{\mu}q+gB^{\mu}j_{\mu}^1+
eA^{\mu}j_{\mu}\end{equation}
where $g$ and $e$ are the strong and electromagnetic couplings,
respectively, and
\begin{equation}
j_{\mu}^1~=~\sum_{q}\bar q\gamma_{\mu}q.\label{current1}
\end{equation}
The sum of electric charges of fermions is taken to be zero to avoid a mixing
between fields $A_\mu$ and $B_\mu$ due to fermionic loops,
\begin{equation}
\sum_{q}Q_q~=~0.
\end{equation}

\indent The simplest renormalon-type graphs are depicted
in Fig.~1 and we will explain the basic features of the technique
on this example.
The sum of these graphs can be cast into the following form:
\begin{equation}
e^2\Pi_{\mu\nu}(q)e^{\mu}_{(1)}e^{\nu}_{(2)}~=~-\frac{1}{2}\int {
{{d^4k}\over{(2\pi)^4}}{{g^2(k^2)}\over {k^2}}
\langle\gamma^*|T|\gamma^*\rangle }\label{ttt}
\end{equation}
where we have made the  Euclidean rotation in the integration over a gluon
momentum $k_\mu$ and
have introduced polarization vectors $e^{\mu}_{(1,2)}$ for initial and
final virtual photons. The running coupling $g^2(k^2)$ sums up vacuum bubble
insertions and has the form
\begin{equation}
g^2(k^2)~=~g^2(Q^2)\left(1+b_0{{g^2(Q^2)}\over {4\pi}}
\ln{{k^2}\over {Q^2}}\right)^{-1}
\label{sss}
\end{equation}
with the first $\beta$ function coefficient
\begin{equation}
b_0~=~-\,{N_f\over {3\pi}}~~ .\label{beta}
\end{equation}
The matrix element $\langle\gamma^*|T|\gamma^*\rangle $ represents the
forward amplitude of gluon-photon scattering and the operator $T$ is
\begin{equation}
T~=~\int dxe^{ikx}T\{j_{\mu}^1(x),j^{1\mu}(0)\}\;.
\label{opT}
\end{equation}
The relation~(\ref{ttt}) is evident for the graphs of Fig.~1. We will
discuss in Sec. 4 how it is modified for higher loops.

By assumption -- to be checked a posteriori --
the momentum $k$ flowing through the gluon line is much larger
than external momentum $Q$. Then it is logical to start by expanding in
inverse powers of $k$. Thus we come to consider
OPE for T-product of two gluonic currents $j_{\mu}^1$
\begin{equation}
T~=~\int{dx~
e^{ikx}~T\{j^1_{\mu}(x)j^{\mu1}(0)\}}~=~\sum
{c_i(k)O_i(0)},\label{oper}\end{equation}
where $O_i$ are local operators. We will discuss this operator expansion in
more detail in the next section.
Here we find coefficients $c_i$ in the tree approximation using the
Schwinger background field technique \cite{Schwinger} (for a review,
see Ref.~\cite{fort}).
In this technique, the lines in the graph of Fig.~2  are understood as
propagators in external electromagnetic and gluonic fields and
the operator $T$ takes the form:
\begin{equation}
T~=~-\sum_{q}\int{dx
\langle x|\bar{q}(X)\gamma^{\mu}{1\over {
{P\!\!\!\!/}+{k\!\!\!/}}}\gamma_{\mu}q(X)|0\rangle +( k \rightarrow -k)}\;.
\label{mel}\end{equation}
Note that within the framework considered operators of the
coordinate $X_{\mu}$ and of momentum $p_{\mu}$ are introduced. Moreover
\begin{equation}
[X_{\mu},p_{\nu}]~=~-ig_{\mu\nu}
,~~[X_{\mu},X_{\nu}]=0,~~[p_{\mu},p_{\nu}]=0
\end{equation}
The operator $P_{\mu}$ is defined as
\begin{equation}
P_{\mu}=p_{\mu}+gB_{\mu}+eA_{\mu}Q_q
\end{equation}
and the matrix element (\ref{mel})
is taken over eigenstates of operator $X_{\mu}$,
\begin{equation}
X_{\mu}|x\rangle ~=~x_{\mu}|x\rangle
.\end{equation}
The construction of the OPE reduces to an expansion of Eq.
(\ref{mel}) in powers of $P_{\mu}$ . The first and second terms of
the expansion
vanish upon averaging over directions of 4-vector $k_{\mu}$ and use of
equations of motion, $iD\!\!\!\!/ \,q=0$.
Thus, the expansion starts from operators
of dimension six,
\begin{equation}
T~=~- {4\over {3k^4}}\sum_{q}\bar{q}\gamma^{\mu}[P^{\nu},[P_{\nu},
P_{\mu}]]q~+~O(k^{-6})\label{rrr}
.\end{equation}
Substituting the commutator
\begin{equation}
[P_{\mu},P_{\nu}]~=~igG_{\mu\nu}+ieF_{\mu\nu}Q_q
\end{equation}
we get
\begin{equation}
T~=~{4\over{3k^4}}\left( e\partial^{\nu}F_{\nu\mu}\sum_{q}Q_q
\bar{q}\gamma^{\mu}q+
gD^{\nu}G_{\nu\mu}\sum_{q}\bar q\gamma^{\mu}q~\right)+~O(k^{-6}).
\label{ope}\end{equation}

The next step is to evaluate the matrix element
$\langle \gamma^*|T|\gamma^*\rangle $. The part of $T$ containing
$D^{\nu}G_{\nu\mu}$ will contribute
only on three-loop level and will be considered in the next section.
As for the part of $T$ containing $\partial^{\nu}F_{\nu\mu}$
it immediately factorizes into
\begin{equation}
\langle \gamma^*|T|\gamma^*\rangle ={4\over {3k^4}}
\left[\, \langle \gamma^*|e\partial^{\nu}F_{\nu\mu}|0\rangle
\langle 0|j^{\mu}|\gamma^*\rangle +
\langle \gamma^*|j^{\mu}|0\rangle \langle
0|e\partial^{\nu}F_{\nu\mu}
|\gamma^*\rangle \, \right]\label{fact}
\end{equation}
where $j^{\mu}$ is the electromagnetic current (see Eq. (\ref{current})).
The matrix element of $\partial^{\nu}F_{\nu\mu}$ is trivial:
\begin{equation}
\langle 0|e\partial^{\nu}F_{\nu\mu}|\gamma^*\rangle =
-e\left(q^2e^{(1)}_{\mu}-q_{\mu}(qe^{(1)})
\right).\label{result}\end{equation}
The matrix element $\langle \gamma^*|j_{\mu}|0\rangle $
is given by the well-known one-loop graph (see Fig.~3) and is equal to
\begin{equation}
\langle \gamma^*|j_{\mu}|0\rangle =
-{{4eN_f\langle Q^2_q\rangle}\over 3}(q^2e^{(2)}_{\mu}-q_{\mu}(qe^{(2)}
))\int{{d^4p}\over{(2\pi)^4p^4}}=
-{{eN_f\langle Q_q^2\rangle}\over{12\pi^2}}\ln{{k^2}\over{Q^2}}\cdot
(q^2e^{(2)}_{\mu}-q_{\mu}(qe^{(2)}))
\label{photon}
\end{equation}
where $\langle Q^2_q\rangle $ is the averaged square of electric charge,
\begin{equation}
\langle Q^2_q \rangle ~=~{1\over {N_f}}\sum_{q}Q^2_q.\end{equation}
Note that the integral over the fermionic loop has
been evaluated with logarithmic accuracy. The upper limit of integration,
$p^2\sim k^2$, is implied by our OPE construction
while the lower bound, $p^2\sim Q^2$, arises
from account in the integrand for the external momentum $Q$.

Substituting the result (\ref{fact}) for
$\langle \gamma^*|T|\gamma^*\rangle $
into (\ref{ttt}) we come to
\begin{equation}
\Pi(Q^2)~=~-{{N_f\langle Q^2_q\rangle }\over{144\pi^4}}Q^2\int_{k^2\sim Q^2}
^{\infty}{{{dk^2}\over{k^4}}\ln{{k^2}\over {Q^2}}\cdot g^2(k^2)}
\label{kk}\end{equation}
Here $Q^2=-q^2$ and the integration over Euclidean $k^2$
runs over $k^2 >Q^2$. We are interested in the expansion of $\Pi(Q^2)$
in $\alpha_1(Q^2)=g^2(Q^2)/4\pi$. Expanding Eq. (\ref{sss}) we get
\begin{equation}
g^2(k^2)~=~4\pi\alpha_1(Q^2)\sum_{n=0}^{\infty}{
\left(-b_0\alpha_1(Q^2)
\right)^n\ln^n{{k^2}\over {Q^2}}}\label{running}\end{equation}
where $b_0$ is the first coefficient in the $\beta$ function.
Performing the integration over $k^2$ in the right-hand side of
Eq.~(\ref{kk}),
\begin{equation}
Q^2\int_{k^2\sim Q^2}^{\infty}{{{dk^2}\over {k^4}}\ln^n{{k^2}\over{ Q^2}}}
{}~=~n!~, \label{prop}\end{equation}
we arrive at the final expression for the UV renormalon contribution:
\begin{equation}
\Pi(Q^2)~=~{{N_f\langle Q^2_q \rangle}\over{36\pi^3b_0}}\sum_{n=1}^{\infty}
\left(-b_0\alpha_1(Q^2)\right)^n~n!~~~. \end{equation}
Differentiating this expression we get
\begin{equation}
-Q^2{{d\Pi(Q^2)}\over {dQ^2}}~=~{{N_f\langle Q_q^2\rangle }\over {12\pi^2}}
\left(-{1\over {3\pi b_0}}\right)
\sum_{n=2}^{\infty}{{n-1}\over n}n!(-b_0\alpha_1)^n.
\end{equation}
In other words, at large $n$ the coefficients
in $\alpha_1$ expansion (for the definition see Eq. (\ref{first}) ) are
given by
\begin{equation}
a_n\stackrel{n\rightarrow\infty}{\longrightarrow}-{1\over {3\pi b_0}}
(-b_0)^n\cdot n!~~~.\label{2loop}\end{equation}
This result coincides with explicit calculations of Ref. \cite{Beneke2}.

To summarize, in this section we utilized the operator product expansion to
evaluate the graphs in Fig.~1 and compared the results with direct
calculations.
The advantage of the operator expansion is not only the compactness of the
calculation but also ensuring automatically the gauge invariance
so that the generalization to QCD case is straightforward and reduces
to the change in $b_0$.
Indeed since the operator product expansion is based on a set of gauge
invariant operators the only dependence on $\ln k^2/Q^2$ arises
through the use of Eq.~(\ref{running}) (or its two-loop generalization)
and we do not need even to specify the gauge fixing
or the class of graphs involved explicitly.
This technical point is especially important in case of non-abelian gauge
theories.

\section{Comments on operator product expansion.}

In this section we combine a few simple remarks on the use of OPE
exemplified in the preceding section.

First, let us demonstrate that Eq.~(\ref{ttt}) which relates the
polarization operator $\Pi_{\mu \nu}$ and the forward amplitude of
gluon-photon scattering is valid in any order of perturbation theory. The
functional dependence on the gluon propagator $D_{\alpha \beta}(x)$ is given
by the following textbook formula:
\begin{equation}
exp\left\{i \int d^4 x {\cal L}_{int} \left(\frac
{\delta}{\delta \eta_\gamma (x)}\right)\right\}
\; \left. exp \left\{ - \int d^4 x d^4 y \,\eta_\alpha (x) D_{\alpha
\beta}(x-y)\eta_\beta (y)\right\}\right|_{\eta_\gamma (x)=0}
\label{Ddep}
\end{equation}
where ${\cal L}_{int}$ is the interaction Lagrangian in which the vector
field $A_\gamma (x)$ is substituted by the functional derivative
$\delta/ \delta \eta_\gamma (x)$ over the source $\eta_\gamma (x)$. The source
$\eta_\gamma (x)$ is substituted by zero after the differentiation. To separate
out the ``hard" part of the gluon propagator $D_{\alpha \beta}^k (x)$ which
corresponds to the exchange by momentum $k$ let us present $D_{\alpha
\beta}(x)$ as
\begin{equation}
D_{\alpha \beta}(x)= D_{\alpha \beta}^k (x)+D_{\alpha \beta}^{soft}(x)\;.
\end{equation}
Then expanding Eq.~(\ref{Ddep}) to the fist order in $D_{\alpha \beta}^k
(x)$ we are proving the relation (\ref{ttt}) in higher orders. What is
crucial for the proof it is the ordering of momenta with the momentum $k$ as
the highest one.

The first comment is to note that our calculation
justifies the assumption  about dominance of large $k^2$. Indeed, the
integrals (\ref{prop})  over $k^2$ are saturated at large $n$ by a saddle
point at
$k^2_{eff}=Q^2e^n$ (see Eq. (\ref{p})).
The width of the range of $k^2$ which contribute to the saddle
point integration is
\begin{equation}
{{k^2-k^2_{eff}}\over{k^2_{eff}}}~\sim~{1\over {\sqrt n}}
\end{equation}
It means, in particular, that the precise value of the lower limit in
integrals (\ref{kk}), (\ref{prop}) is of no importance at large $n$.

Thus, expansion in $Q^2/k^2$ is fully justified. Moreover, it is clear
from the
calculations above that it is the dimension of operators $O_i$ (see Eq.
(\ref{oper})) which is most important. Namely, for an operator of
dimension $d$ the contribution to $a_n$ is proportional to
\begin{equation}
Q^{d-4}\int_{k^2\sim Q^2}^{\infty}{
{{dk^2}\over {k^{d-2}}}\ln^n{{k^2}\over {Q^2}}~=~{{n!}\over
{\left((d-2)/2\right)^{n+1}}}}
.\end{equation}
In the preceding section we dealt with operators of dimension $d=6$.
Operators of higher dimensions give rise contributions which are
suppressed by powers of $(d-4)/2$ and can be safely neglected.

However, the question naturally arises on the role of operators of
dimension four. Although such operators did not appear in the tree
approximation of the preceding section they
show up via loop corrections. It is worth emphasizing therefore that these
operators are not relevant to asymptotic of $a_n$.

Indeed, matrix elements of $d=4$ operators over virtual photons have the
form
\begin{equation}
\langle \gamma^*|O_i^{d=4}|\gamma^*\rangle~=~A_ie^2\left(q^2(e^{(1)}e^{(2)})
- (qe^{(1)})(qe^{(2)})\right)
\end{equation}
where $A_i$ are dimensionless and the corresponding contributions to
$\Pi (Q^2)$ are \begin{equation}
\Delta_i\Pi~\sim~\int dk^2A_ic_i^{d=4}(k^2)g^2(k^2)
\label{uv}
\end{equation}
where $c_i^{d=4}$ are OPE coefficients, $c_i^{d=4}\sim1/k^2$. To get rid
of the apparent  ultraviolet divergence in (\ref{uv}) one needs to consider
instead of $\Pi(Q^2)$ the quantity $Q^2\cdot d\Pi(Q^2)/dQ^2$ which contains
no ultraviolet uncertainty.

The finiteness of $Q^2\cdot d\Pi(Q^2)/dQ^2$ implies that
$\sum_i c_i^{d=4} A_i$ is independent of $Q^2$. Notice that this assertion
is valid in spite of nonzero anomalous dimensions of some $d=4$ operators.
It also persists in the presence of fermionic mass terms $m_q \bar q q$
which contains $d=3$ operators. The $Q^2$ independence $\sum_i c_i^{d=4}
A_i$ in any order of perturbation theory means that the sum of
contributions (\ref{uv}) drops off from $Q^2\cdot d\Pi(Q^2)/dQ^2$.

Thus $d=4$ operators do not contribute to UV renormalons and the UV
renormalon calculus deals with $d=6$ operators. In the preceding section
we had  one such operator explicitly:
\begin{equation}
O_{Fj}~=~e\partial^{\nu} F_{\nu\mu}j^{\mu}
\label{one}
\end{equation}
where $j_{\mu}$ is defined by Eq. (\ref{current}). The calculation of matrix
element $\langle \gamma^*|O_{Fj}|\gamma^*\rangle $ (see Eq. (\ref{fact}))
can be interpreted as a mixing of $O_{Fj}$ with operator
\begin{equation}
O_{F^2}~=~\left( e\partial^{\nu}F_{\nu\mu}\right)^2\label{four}
,\end{equation}
i.e. with logarithmic accuracy we have
\begin{equation}
O_{Fj}|_{k^2}~=~O_{Fj}|_{Q^2}+{{N_f\langle Q_q^2\rangle }\over {12\pi^2}}
 \ln{{k^2}\over {Q^2}}\cdot O_{F^2}|_{Q^2}\label{log}
\label{mixing}\end{equation}
where we marked the normalization points of the operators.
Since the diagonal
anomalous dimensions of the operators $O_{Fj},O_{F^2}$ are vanishing
Eq.~(\ref{mixing}) specifies the matrix of the anomalous dimensions
completely.

Note also that the log in the right-hand side of Eq. (\ref{log}) results in
an enhancement  of $a_n$ by factor of $n$ (see Eq. (\ref{prop})) and this
enhancement is  crucial
for the consistency of our calculation. The point is that in one-loop order
the operator $O_{F^2}$ appears in the OPE (\ref{oper}) not only through its
mixing with $O_{Fj}$ as in the tree approximation but directly as well.
In that
case large momentum $k$ flows through all fermionic lines in graph of Fig.~1.
However then there is no log factor similar to the one in Eq. (\ref{mixing})
so that the direct $O_{F^2}$ coefficient can be neglected with $1/n$ accuracy.
There are also $d=6$ four-fermionic operators which contribute to $\Pi(Q^2)$
on three-loop level and these will be discussed in the next section. Here we
would like only to note that their logarithmic mixing with the operators
$O_{Fj}$ and $O_{F^2}$ results in an enhancement of the three-loop skeleton
graph.

Finally we note that so far we considered for simplicity one-loop
$\beta$ function for $\alpha_1$. Accounting for the second coefficient
in the $\beta$ function brings a non-vanishing constant $\gamma$ in Eq.
(\ref{factorial}). The procedure is rather standard and we will
give examples in the next section. As for the overall normalization
(constant $K$ in Eq. (\ref{factorial})) it is actually scheme dependent,
for further discussion see \cite{Beneke3,Grunberg}.
The constant $K$ can be fixed in large $N_f$ limit, which simplifies
calculations greatly (see, e.g., \cite{Beneke2,Broadhurst,Maxwell}).
As we shall see later, however, the limits of $n$ and $N_f$ tending
to infinity are not necessarily consistent with each other. Here we keep
track of large $n$ dependence.

To summarize, the operator product expansion provides with systematic means
of calculating asymptotic of the perturbative expansion coefficients $a_n$
associated with renormalon-type graphs.

\section{Three-loop example.}

In this section we evaluate the renormalon contribution
associated with the graphs with two chains of vacuum-polarization insertions,
or three-loop skeleton graphs, see examples in Fig.~4.
In fact we have already mentioned that the contribution of
the operator $D^{\nu}G_{\nu\mu}$
(see Eq. (\ref{photon})) arises only on a three-loop level and now we come
to consider three-loop calculations in more detail.

Note that $D^{\nu}G_{\nu\mu}$ can be substituted by $(-g)\sum\bar{q}
\gamma_{\mu}q$ via equations of motion so that we get for the
corresponding part $T_1$ of operator $T$ defined by Eq.~(\ref{opT})
\begin{equation}
T_1~=~-{4\over 3}{{g^2}\over
{k^4}}O_1,~~~~~O_1=
\left(\sum_{q}\bar{q}\gamma_{\mu}q\right)^2\;.
\label{two}\end{equation}
Diagramatically the appearance of $T_1$ in three-loop graphs for the
polarization operator is illustrated by the diagram $b$ in Fig.~4.
The dotted box in this diagram presents a ``penguin" mechanism for $T_1$
similar to the one in weak interactions~\cite{weak}.

A different mechanism leading to four-fermion operators is due to the exchange
by two gluons, see the diagram $a$ in Fig.~4. The part of the graph which
corresponds to the operator $T$ is marked by the dotted box in Fig. 4$a$ and
presented separately in Fig.~5. As far as explicit calculations of the graphs
are concerned they are very  similar to those used in derivation of QCD sum
rules \cite{SVZ1} since OPE is exploited in the both cases. The difference is
that in the former case it is the momentum carried by the electromagnetic
current which is considered to be large while now the large momentum is
brought in by the  gluonic current.

The subtle point here is that the relation between the scattering diagram of
Fig.~5 and the box part of diagram $a$ in Fig.~4 contains a combinatorial
coefficient 1/2 which is absent for the ``penguin" box of Fig.~4$b$.
Indeed, the
functional dependence on the gluon propagator $D_{\alpha \beta}(x)$ is given
by the following textbook formula:
\begin{equation}
exp\left\{i \int d^4 x {\cal L}_{int} \left(\frac
{\delta}{\delta \eta_\gamma (x)}\right)\right\}
\; \left. exp \left\{ - \int d^4 x d^4 y \,\eta_\alpha (x) D_{\alpha
\beta}(x-y)\eta_\beta (y)\right\}\right|_{\eta_\gamma (x)=0}
\label{Ddep}
\end{equation}
where ${\cal L}_{int}$ is the interaction Lagrangian in which the vector
field $A_\gamma (x)$ is substituted by the functional derivative
$\delta/ \delta \eta_\gamma (x)$ over the source $\eta_\gamma (x)$. The source
$\eta_\gamma (x)$ is substituted by zero after the differentiation. To separate
out the ``hard" part of the gluon propagator $D_{\alpha \beta}^k (x)$ which
corresponds to the exchange by momentum $k$ let us present $D_{\alpha
\beta}(x)$ as
\begin{equation}
D_{\alpha \beta}(x)= D_{\alpha \beta}^k (x)+D_{\alpha \beta}^{soft}(x)\;.
\end{equation}
Then expanding Eq.~(\ref{Ddep}) to the fist order in $D_{\alpha \beta}^k
(x)$ we get the relation (\ref{ttt}).  To maintain the same relation
(\ref{ttt}) for terms of the second order in
$D_{\alpha \beta}^k (x)$ the corresponding part of gluon scattering
amplitude should be taken with the extra factor $1/2\,!$.

Accounting for the modification described we get
the following result for four-fermion operators generated  by graphs of the
type given by Fig.~5:
\begin{equation}
T_2~=~-{{3 g^2}\over {k^4}}O_2,~~~~~O_2=
\left(\sum_{q}\bar{q}\gamma_{\mu}\gamma_5 q\right)^2\label{three}\;.
\end{equation}
Dotted boxes in Fig.~4 present are reduced to local operators for fermionic
momenta much less than $k$. We also account for the graphs which gives
logarithmical dressing of the boxes, i.e.
$(\alpha
\ln (k^2/Q^2))^n$ corrections. These corrections are due to the range of
momenta
between
$k$ and $Q$ and diagrammatically are graphs presenting gluon exchanges between
 legs of effective four-fermionic operators as well as penguin-type fermionic
loops for these operators. They were absent in two-loop considerations because
of vanishing anomalous dimension of the relevant operator $O_{Fj}$ (see Eq.
(\ref{one})).  Operators $O_{1,2}$ have nonzero anomalous dimension.

To sum up logarithmic corrections for the matrix element $\langle \gamma
^*|T|\gamma^*\rangle $ we apply the standard machinery of
the renormalization group.
The operator basis of the problem consists of operators
$O_1,~O_2,~O_{Fj},~O_{F^2}$
(see Eqs. (\ref{two},~\ref{three},~\ref{one},~\ref{four})).
Their coefficients $c_1,~c_2,~c_{Fj},~c_{F^2}$ are functions of
$\mu^2$ where $\mu$ is the normalization point.

The set of renormalization group equations in one-loop approximation looks as
\begin{eqnarray}
\mu^2{d\over {d\mu^2}}\,c_{F^2}& = &-{{N_f \langle Q_q^2\rangle}
\over{12\pi^2}}c_{Fj}\;,\nonumber\\
\mu^2{d\over {d\mu^2}}\,c_{Fj}&=&-{1\over {12\pi^2}}(c_1+c_2)\;,\\
\label{set}
\mu^2{d\over{d\mu^2}}\,c_1&=&\frac{\alpha_1}{\pi}
\left(\frac{2N_f+1}{3}c_1 + \frac{11}{6} c_2 \right)\;,\nonumber\\
\mu^2{d\over{d\mu^2}}\,c_2&=&\frac{3\alpha_1}{2\pi}c_1\;.\nonumber
\end{eqnarray}
The solution for $c_1$, $c_2$ has the form
\begin{eqnarray}
c_1(\mu^2) & = & \frac{1}{1+(11/4\pi^2 b_0^2\gamma_1^2)} \left\{\left[
c_1(\mu_0^2)+
\frac{11}{6\pi b_0 \gamma_1} c_2(\mu_0^2) \right] \left[
\frac{\alpha_1(\mu_0^2)}{\alpha_1(\mu^2)} \right]^{\gamma_1}+
\right. \nonumber\\
 & & \left. \left[ \frac{11}{4\pi^2 b_0^2 \gamma_1^2} c_1(\mu_0^2)-
\frac{11}{6\pi b_0 \gamma_1}
c_2(\mu_0^2) \right] \left[
\frac{\alpha_1(\mu_0^2)}{\alpha_1(\mu^2)} \right]^{\gamma_2}\right\} \\
c_2(\mu^2)&=&\frac{1}{1+(11/4\pi^2 b_0^2 \gamma_1^2)} \left\{\left[
\frac{3}{2\pi b_0 \gamma_1} c_1(\mu_0^2) +\frac{11}{4\pi^2 b_0^2\gamma_1^2}
c_2(\mu_0^2)\right] \left[
\frac{\alpha_1(\mu_0^2)}{\alpha_1(\mu^2)} \right]^{\gamma_1} +
\right. \nonumber\\
 & & \left. \left[ -\frac{3}{2\pi b_0 \gamma_1} c_1(\mu_0^2)
+c_2(\mu_0^2)\right]
\left[ \frac{\alpha_1(\mu_0^2)}{\alpha_1(\mu^2)} \right]^{\gamma_2}\right\}\;,
\end{eqnarray}
where anomalous dimensions $\gamma_{1,2}$ are
\begin{equation}
\gamma_{1,2}=\frac{1}{\pi b_0}\,\left( \frac{2N_f+1}{6} \pm
\sqrt{\left(\frac{2N_f+1}{6}\right)^2+\frac {11}{4}}\;\right)\;.
\label{dim}
\end{equation}
Coefficients $c_{Fj}$ and $c_{F^2}$ are obtained then by a simple
integration.
Initial conditions for $c_i(\mu^2)$ are set by Eqs. (\ref{two},
\ref{three}) at $\mu^2=k^2$,
\begin{equation}
c_1(k^2)=-\frac{4}{3}\;\frac{g^2(k^2)}{k^4}\;,\;\;
c_2(k^2)=-3\; \frac{g^2(k^2)}{k^4}\;,\;\;
c_{Fj}(k^2)=c_{F^2}(k^2)=0\;.
\end{equation}

We need to calculate the value of $c_{F^2}(\mu^2=Q^2)$ because
with logarithmic accuracy
the matrix element $\langle \gamma^*|T|\gamma^*\rangle $ is given by
\begin{equation}
\langle\gamma^*|T|\gamma^*\rangle~=~2\, e^2 c_{F^2}(\mu^2=Q^2)\,q^2
\left[ q^2(e^{(1)}e^{(2)})-(qe^{(1)})(qe^{(2)})\right]
\end{equation}
The result for $c_{F^2}(\mu^2=Q^2)$ is rather lengthy expression,
$$c_{F^2}(\mu^2=Q^2)=-\frac{1}{9} N_f \langle Q_q^2\rangle\,\frac{1}{k^4}\,
\frac{1}{g^2(k^2)}\,\frac{1}{\pi^2 b_0^2 + (11/4\gamma_1^2)}\times$$
$$\left\{ \left(\frac{4}{3} + \frac{15}{2\pi b_0\gamma_1}+
\frac{33}{4\pi^2 b_0^2\gamma_1^2}\right)\frac{1}{1+\gamma_1}
\left[1-\kappa - \frac{1}{2+\gamma_1}\left(1-\kappa^{(2+\gamma_1)}\right)
\right]\right.
$$
\begin{equation}
+\left. \left(3-\frac{15}{2\pi b_0\gamma_1}+
\frac{11}{3\pi^2 b_0^2\gamma_1^2}\right)\frac{1}{1+\gamma_2}
\left[1-\kappa -
\frac{1}{2+\gamma_2}\left(1-\kappa^{(2+\gamma_2)}\right)\right]\right\}\;
\end{equation}
where
\begin{equation}
\kappa=\frac{\alpha_1(k^2)}{\alpha_1(Q^2)}=\left[1+b_0\alpha_1(Q^2)\ln
\frac{k^2}{Q^2}\right]^{-1}\;.
\end{equation}
The polarization operator $\Pi$ is given then by integration
over $k$,
\begin{equation}
\Pi~=~-\frac{Q^2}{16\pi^2}\int_{Q^2}^\infty dk^2
\,g^2(k^2)\,c_{F^2}(\mu^2=Q^2)\;.
\end{equation}
Coefficients of expansion in $\alpha_1$ are defined by integrals of the
type
\begin{equation}
Q^2\int_{Q^2}^{\infty}
\frac{dk^2}{k^4}\left[\alpha_1(k^2)\right]^\delta~=~
\left[\alpha_1(Q^2)\right]^\delta\,{\sum_{n=0}^{\infty}}
\frac{\Gamma(n+\delta+\delta_1)}{\Gamma(\delta+\delta_1)}
\left[-b_0\alpha_1(Q^2)\right]^n
\label{anom}
\end{equation}
where $\delta_1$ arises from account of the second coefficient in
the $\beta$ function $b_1$ (see, e.g., \cite{mueller}) and equals to
\begin{equation}
\delta_1~=~-{{b_1}\over{ b^2_0}}\;.
 \end{equation}

The final formula for the expansion of $\Pi$ in powers of $\alpha_1(Q^2)$
has the form:
$$
\Pi~=~\frac{N_f \langle Q_q^2\rangle}{144\pi^2}\,
\frac{1}{\pi^2 b_0^2 + (11/4\gamma_1^2)}\sum_{n=2}^{\infty}
\left[-b_0\alpha_1(Q^2)\right]^n\times$$
$$
\left\{ \left(\frac{4}{3} + \frac{15}{2\pi b_0\gamma_1}+
\frac{33}{4\pi^2 b_0^2\gamma_1^2}\right)\frac{1}{1+\gamma_1}
\left[\frac{\Gamma(n+2+\gamma_1+\delta_1)}{\Gamma(3+\gamma_1+\delta_1)}
- \Gamma(n+1+\delta_1) \right]\right.+
$$
\begin{equation}
\left. \left(3- \frac{15}{2\pi b_0\gamma_1}+
\frac{11}{3\pi^2 b_0^2\gamma_1^2}\right)\frac{1}{1+\gamma_2}
\left[\frac{\Gamma(n+2+\gamma_2+\delta_1)}{\Gamma(3+\gamma_2+\delta_1)}
-\Gamma(n+1+\delta_1)\right]\right\}
\label{pifin}
\end{equation}

Assuming for the moment
$\gamma_1=\gamma_2=\delta_1 =0$, i.e. omitting effects of anomalous
dimensions and two-loop $\beta$ function, we get for coefficients $a_n$
defined by Eq. (\ref{first})
\begin{equation}
a_n\stackrel{n\rightarrow \infty}{\longrightarrow}~=~ -{13 \over
{36\pi^2 b_0^2}}
\left( -b_0\right)^n(n+1)!
\end{equation}
Comparing this result with the Eq. (\ref{2loop}) we see
that the contribution of the four-fermion
operators, or of the three-loop graphs, dominates at large $n$ over that
of the two-loop graph, it contains an extra factor $n$.
Moreover, this conclusion is not modified if
nonzero $\gamma_i$ and $\delta_1$ are accounted for. Indeed, the dependence
on $\delta_1$ is universal for two-loop and three-loop contributions and
drops off from the ratio of $a_n$.
As for the anomalous dimensions $\gamma_i$ of the four-fermionic operators
(see Eq.~(\ref{dim})), $\gamma_1$ is negative, $\gamma_2$ is positive.
Notice that
in the realistic case of QCD one also gets quite a few different diagonal
operators both with positive and negative anomalous
dimensions~\cite{SVZ1,broadhurst}.
In the large $n$ limit the operator with the most positive
$\gamma$ dominates so that account for the anomalous dimensions only
strengthens the conclusion on the dominance of three-loop
graphs.

In the particular case when the number of
flavors $N_f$ is large, $N_f \gg 1$, one arrives at $\gamma_1=-2$,
$\gamma_2=0$,
$\delta_1=0$ and Eq.~(\ref{pifin}) leads to
\begin{equation}
a_n\stackrel{n\rightarrow \infty}{\longrightarrow}~=~
-\frac{9}{8 N_f^2}\left( -b_0\right)^n(n+1)!\;\;.
\end{equation}
Although in large $N_f$ limit the three-loop contribution is suppressed by
an extra factor $1/N_f$ as compared with the two-loop one it has an extra
factor $n$.

As an example of advantage of OPE technique used above let us mentioned that
the summation automatically accounts for rather complicated graphs, in
particular, those where the high momentum gluon propagator  is inside of
fermionic bubble for a low ($\sim Q$) momentum gluon.

To summarize, we have demonstrated in this section that three-loop graphs
generate new, four-fermionic operators. The technique of the operator product
expansion works the same simple as in the two-loop case. Although the final
expressions accounting for the anomalous dimensions are somewhat
cumbersome they are straightforward. The most non-trivial conclusion is
that four-fermionic operators dominate at large $n$.

\section{Calculability of ultraviolet renormalon. Conclusions.}

In this section we comment on the definition of the ultraviolet renormalon.
The basic observation is
that if one tries to identify the ultraviolet renormalon as a set
of graphs producing a certain asymptotic behavior, $a_n\sim n!(b_0)^n$,
then this set of graphs is in fact ill-defined.

The demonstration of this point is straightforward.
Let us go back to our operator product expansion (\ref{oper}).
The coefficient functions are represented as series in $\alpha_1(k^2)$
\begin{equation}
c_i(k^2)~=~h_0+h_1\alpha_1(k^2)+h_2\alpha^2_1(k^2)+
...+h_l\alpha^l_1(k^2)+...\label{newex}
\end{equation}
The standard and tacit assumption is that one can approximate $c_i(k^2)$ by,
say, first term in this expansion as far as $\alpha_1(k^2)$ is small. This
logic
does not work however if we use the operator product expansion (\ref{oper})
to evaluate the asymptotic of the coefficients $a_n$ of expansion in our
``original'' $\alpha_s(Q^2)$. Indeed, from Eq. (\ref{anom}) we conclude that
for any finite $l$ the contribution of $h_l$ to the asymptotic
of $a_n$ has the same $n$ dependence. This means, in turn, that all the
terms in the expansion (\ref{newex}) are the same important.

Technically this happens because originally there are two large parameters,
$n$ and $ln(k^2/Q^2)$, and only the logs are controlled by the renormalization
group. As a result of integration over $k^2$ powers of logs are converted into
powers of $n$ (see, e.g., Eq. (\ref{prop})) and the two parameters merge. The
reason why all $h_l$ above contribute the same order of magnitude to
asymptotical $a_n$ is that lower powers of logs have larger statistical
weights.
This can be traced directly by inspecting combinatorics of $l$ renormalon
chains of graphs. As a result the two large parameters, combinatorial
$n$ and $\ln k^2$, get mixed up and the actual asymptotic of $a_n$ is
determined
by an extremum in the two-dimensional parameter space. Finding this extremum
looks much harder a problem than fixing the leading logs. It is not ruled out,
in particular, that the contributions of various $h_l$ in Eq. (\ref{newex})
cancel between themselves and the true asymptotic is very different from
(\ref{factorial}), (\ref{UV}).

It is worth emphasizing that this difficulty with obtaining the asymptotic of
the expansion coefficients is specific for the ultraviolet renormalon and does
not plague the calculation of the infrared renormalon. The reason for this
looks pure technical. On the other hand, renormalons are reflections of the
Landau poles in the running couplings (see, e.g., \cite{Book}) and if indeed
the true asymptotic of the UV renormalon is different from
(\ref{factorial}),(\ref{UV}) then the Landau pole in ultraviolet does
not manifest itself in, say, polarization operator $\Pi(Q^2) $.
Thus there is a remarkable difference in the statuses of the infrared and
ultraviolet renormalons.

As another illustration of the uncertainties in the evaluation of the
ultraviolet region in perturbation theory let us consider a tower of
renormalons which means that we introduce another scale, $k'$ such that
\begin{equation}
(k')^2~\gg~k^2~\gg~Q^2,~~~(k')^2~\sim~k^2\, e^l \end{equation}
and $l$ is large enough. Then coefficients $h_l$ are given by
\begin{equation}
h_l~\sim~l\,!\left(-b_0\right)^l\;
.\end{equation}
The corresponding contribution of each $h_l$ to the asymptotic of $a_n$ is
again
\begin{equation}
{a_n}\stackrel{n\rightarrow\infty}{\longrightarrow}~ const\cdot n\,!(-b_0)^n.
\end{equation}
Moreover, we ignored the anomalous dimension of the operators which arise
at the scale $k'$ so that the true asymptotic can again be enhanced.
But even these simple estimates suffice to demonstrate that no matter how
large the virtual momenta are they do not detach from the asymptotic of $a_n$.

Thus, we conclude that at present there are no technical means to evaluate the
true asymptotic of the UV renormalon.

As for possible phenomenological implications
of the results obtained we have unfortunately little to say. Pessimistically,
one may argue that all the problems with evaluating the UV asymptotic are not
physical and the reason why we fail is simply because we want to represent
as a series in $\alpha_s(Q^2)$ the coupling $\alpha_s(k^2)$ which is much
smaller than $\alpha_s(Q^2)$ at $k^2\gg Q^2$. To this end we need to keep
many sign alternating terms absolute value of which is much larger than
$\alpha_s(k^2)$ itself. If we do not endeavor the expansion, however, the
whole UV contribution is small and uninteresting.

On the optimistic side, one might say that the divergences of the
perturbative expansions reveal power like corrections and the failure to
evaluate the UV renormalon signifies that the corresponding $1/Q^2$ terms
are important. Indeed, applying the Borel summation one can convert the UV
renormalon series into $1/Q^2$ term:
\begin{equation}
K\sum_{N~large}^{\infty}\alpha^n_s(Q^2)(-1)^n (b_0)^n n\,!
\longrightarrow {K\over {Q^2}}
\end{equation}
and if the constant $K$ happens to be large numerically then the $Q^{-2}$
correction is also large.
Moreover, one can generate the $Q^{-2}$ terms without invoking the Borel
summation either by varying the normalization point of $\alpha_s$ in the
expansion \cite{Beneke4} or by exploiting the conformal mapping of
perturbative expansions \cite{Mueller1}.

One general remark concerning the power like corrections is now in order.
Generally speaking, one would not expect these corrections be important
numerically since they are naturally screened by lower order perturbative
terms. The examples of successful phenomenological applications are known
so far for IR renormalons only. For example, the first IR renormalon is
associated with $Q^{-4}$ term \cite{mueller}. Phenomenologically, the
corresponding correction is described by the so called gluon condensate
\cite{SVZ1}, $\langle 0|\alpha_s(G_{\mu\nu}^a)^2| 0\rangle $ where
$G_{\mu\nu}^a$
is the gluonic field strength tensor. The basic assumption is that this
condensate is large and exceeds at moderate $Q^2$ even the lowest terms
of the perturbative expansion. Having only perturbative expansions in hand,
this assumption is very difficult to justify.

Similarly, to have the UV renormalon important we need to assume that
the constant $K$ is large. Better to say, we need a mismatch between
large $K$ in the asymptotic and numerical values of the lowest $a_n$.
The complicated structure of the UV renormalon revealed above may be
considered as suggesting such a complicated structure of the perturbative
series. The main problem is not that such an assumption would be too bizarre
but rather lack of a framework which would allow to develop a phenomenology
basing on this assumption. Indeed, in case of the $Q^{-4}$ terms the main
point was the possibility to relate these terms in various channels, that
is not possible at present in case of the UV renormalon.

Although we do not have constructive ways to introduce this phenomenology let
us note that the natural direction would be Nambu-Jona-Lasinio type models
\cite
{nambu} since in these models one exploits four-fermion interaction as a
low-energy effective interaction in QCD. We have seen that UV renormalon is
naturally related to four-fermion interaction as well. One could hope to match
$1/Q^2$ corrections due to the renormalon with NJL type phenomenology.

\section{Acknowledgments}

The authors are grateful to M. Beneke, V. Gribov, G. Grunberg and
M. Shifman for useful discussions. We are particularly thankful to M.~Beneke,
the comparison with his unpublished results helped us to localize some mistakes
in the preprint version of the paper. This work was supported in part by  DOE
under the grant number DE-FG02-94ER40823.

\newpage
{\bf Figure captions.}\\

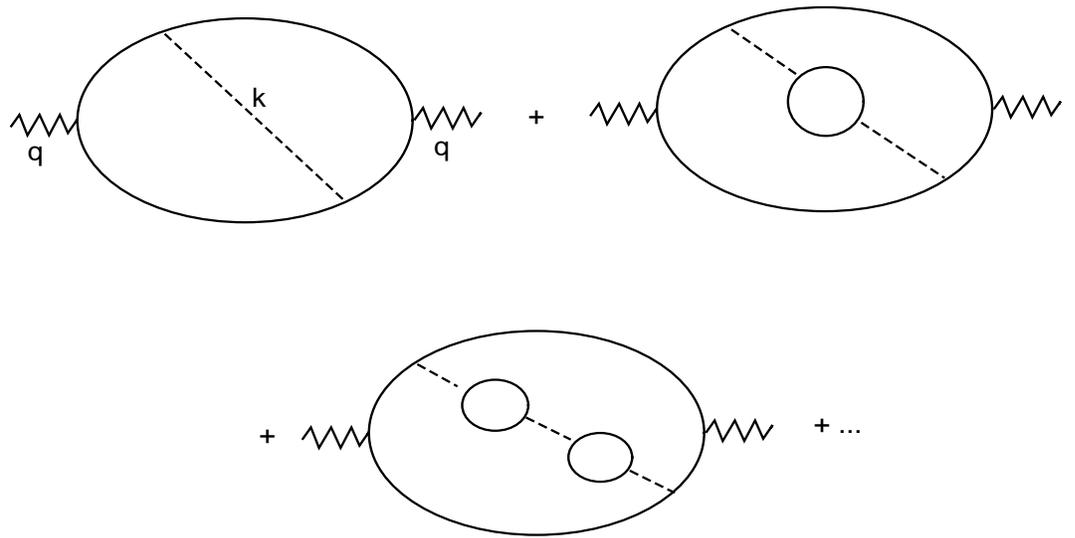
{\bf Fig 1.} Building up the simplest renormalon-type graph. Dashed line
denotes
gluon while solid lines refer to fermions. One starts with an
exchange of a vector particle of momentum $k$ and inserts vacuum polarization
bubbles n times.\\

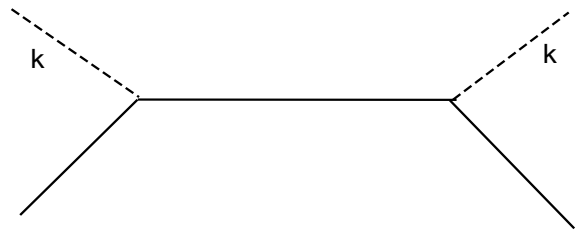
{\bf Fig 2.} The graph used to evaluate coefficients of operator expansion
associated
with the simplest renormalon graphs of Fig. 1. Momentum $k$ carried by the
gluonic line is considered to be large. The fermion is understood to
propagate in external electromagnetic and gluonic fields so that the graph
is in fact a subgraph of the one-gluon exchange depicted in Fig. 1. \\

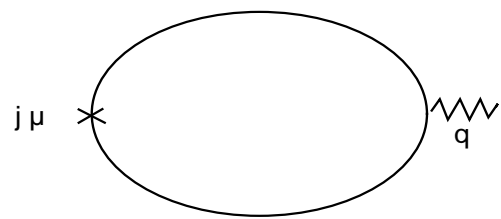
{\bf Fig 3.} One-loop graph describing transition of electromagnetic current
$j_{\mu}$ into a virtual photon with momentum $q$.\\

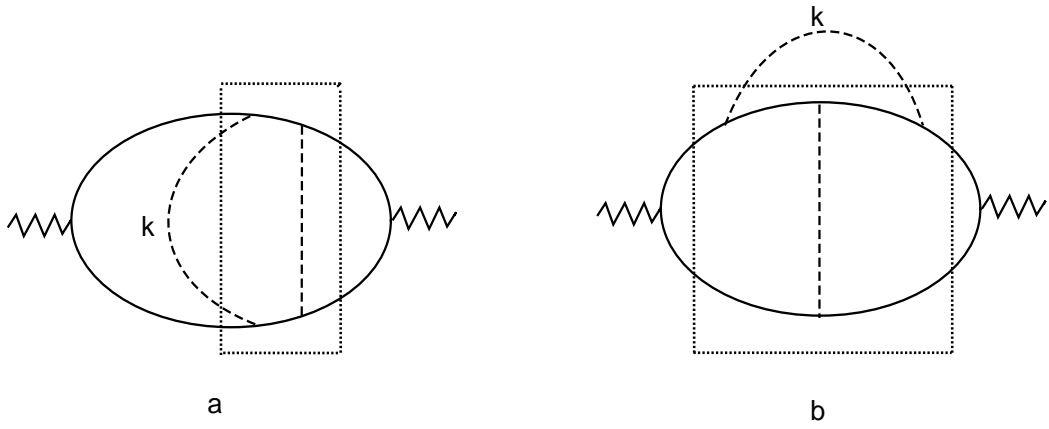
{\bf Fig 4.} Three-loop skeleton graphs giving rise to four-fermionic
operators.
Momentum $k$ flowing through the gluonic lines is considered to be large
so that the operator product expansion is an expansion in inverse powers of
$k^{-2}$. The dotted boxes mark subgraphs producing four fermionic operators
$O_{2}$ and $O_{1}$ (see Eqs. (\ref{three}) and (\ref{two}), respectively). \\

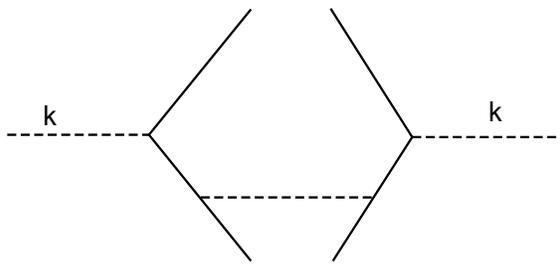
{\bf Fig 5.} The graph giving rise to the operator $O_{2}$ in the limit
of large $k$. It is a subgraph of the first graph in Fig.~4 where it is
marked by a dotted box.

\end{document}